\documentclass[10pt,aps,prl,twocolumn,notitlepage,bibnotes,longbibliography,
floatfix,showpacs,citeautoscript,superscriptaddress]{revtex4-1}

\usepackage[english]{babel}	

\usepackage{amsmath}		
\usepackage{amssymb}		
\usepackage{bm}

\usepackage{graphicx}		
\usepackage{graphics}		
\usepackage{color}		
\usepackage{xcolor}		
\DeclareGraphicsExtensions{.png,.jpg,.pdf,.eps}	

\newcommand\be{\begin{eqnarray}}
\newcommand\ee{\end{eqnarray}}
\newcommand{\bs}{\boldsymbol}
\newcommand{\eq}[1]{(\ref{#1})}
\newcommand{\ch}[2]{#1${}_{#2}$}

\newcommand{\Figref}[1]{Fig.~\ref{#1}}
\newcommand{\Eqref}[1]{\eqref{#1}}

\newcommand{\Grad}{\bs \nabla}

\newcommand{\Curl}{\bs \nabla\times}

\newcommand{\D}{{\bs D}}

\newcommand{\A}{{\bs A}}

\newcommand{\J}{{\bs J}}
\renewcommand{\j}{{\bs j}}
\newcommand{\h}{{\bs h}}

\begin{document}

\title{Chiral Magnetic Josephson junction:\\ a base for low-noise superconducting qubits?}

\author{M. N. Chernodub}
\email{maxim.chernodub@idpoisson.fr}
\affiliation{Institut Denis Poisson CNRS/UMR 7013, Universit\'e de Tours, 37200 France}
\affiliation{Laboratory of Physics of Living Matter, Far Eastern Federal University, Sukhanova 8, Vladivostok, 690950, Russia}
\
\author{J. Garaud}
\email{garaud.phys@gmail.com}
\affiliation{Institut Denis Poisson CNRS/UMR 7013, Universit\'e de Tours, 37200 France}
\
\author{D. E. Kharzeev}
\email{dmitri.kharzeev@stonybrook.edu}
\affiliation{Department of Physics and Astronomy, Stony Brook University, New York 11794-3800, USA}
\affiliation{Department of Physics and RIKEN-BNL Research Center,\\ Brookhaven National Laboratory,  Upton, New York 11973, USA}
\affiliation{Le Studium, Loire Valley Institute for Advanced Studies, Tours and Orl\'eans, France}

\date{\today}

\begin{abstract}

Superconducting materials with non-centrosymmetric lattices lacking the space inversion symmetry are known to exhibit a variety of interesting parity-breaking phenomena, including the anomalous Josephson effect. Here we consider a Josephson junction consisting of two non-centrosymmetric superconductors (NCSs) connected by a uniaxial ferromagnet, and demonstrate that it exhibits a direct analog of the Chiral Magnetic Effect observed in Dirac and Weyl semimetals. We propose to use this ``Chiral Magnetic Josephson junction" (CMJ junction) as an element of a qubit with a Hamiltonian tunable by the ferromagnet's magnetization. The CMJ junction allows to avoid the use of an offset magnetic flux in inductively shunted qubits, thus enabling a simpler and more robust architecture. 
The resulting ``chiral magnetic qubit" is protected from the noise caused by fluctuations in magnetization when the easy axis of the uniaxial ferromagnet is directed across the junction.
\end{abstract}

\maketitle

The discovery of superconductors lacking the spatial inversion symmetry \cite{Bulaevskii.Guseinov.ea:76,Levitov.Nazarov.ea:85,CePt,Samokhin,Sigrist} has opened the possibility to study spontaneous breaking of a continuous symmetry in a parity-violating material. In particular, the superconducting order parameter in these non-centrosymmetric superconductors (NCSs) is a parity--odd quantity~\cite{Samokhin,Sigrist}, enabling a number of interesting magnetoelectric phenomena due to mixing of singlet and triplet superconducting parameters, correlations between supercurrents and spin polarization, appearance of helical states and peculiar structure of Abrikosov vortices (see \cite{Bauer.Sigrist,NCS1} for a review).

\vskip0.3cm
Parity breaking in NCSs also results in an unconventional Josephson effect, where the junction features a phase-shifted current relation \cite{B2008,KB2009}:
\be\label{current}
J(\varphi, \varphi_g) = J_c \sin(\varphi - \varphi_g).
\ee
Here $\varphi$ is the superconducting phase difference across the junction, $J_c$ the critical Josephson current, and $\varphi_g$ is the parity-breaking phase offset. 
Nonzero bias $\varphi_g \neq 0$ results in a nonvanishing current across the junction, even when the phase difference $\varphi$ is zero. Since the current is a parity-odd quantity, this clearly signals parity violation.
\vskip0.3cm

Phase-biased junctions (often called ``$\varphi_0$--junc\-tions'') were suggested to appear in a wide range of systems including non-centrosymmetric~\cite{B2008}, and multilayered~\cite{Liu:2010} ferromagnetic links between conventional superconductors, topological insulators~\cite{ref:top:1,ref:top:2}, nanowires~\cite{ref:nanowires:1,ref:nanowires:2}, quantum point contacts~\cite{Reynoso}, and quantum dots~\cite{ref:dot:1,ref:dot:2,ref:dot:3}. The first experimental realization of Josephson $\varphi_0$-junctions has been reported in superconductor--quantum dot structures, where the phase offset $\varphi_g$ can be controlled via electrostatic gating~\cite{ref:phi0:experiment}.

\begin{figure}[!thb]
\hbox to \linewidth{ \hss
\includegraphics[width=0.7\linewidth]{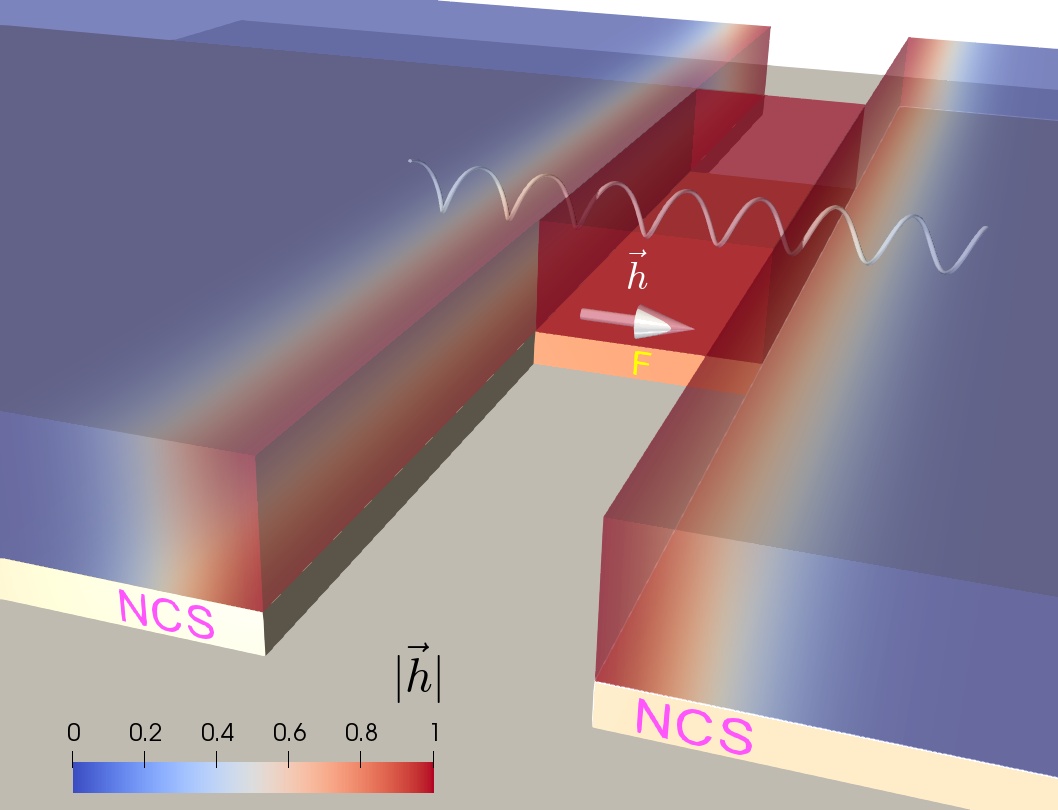}
\hss}
\caption{
The Chiral Magnetic Josephson junction: two non-\-centrosymmetric superconductors (NCSs) weakly linked by a uniaxial ferromagnet (F). The exchange field $\h$ of the ferromagnet, oriented across the link, induces an inversion symmetry-breaking component of the supercurrent (represented here by the spiral) in the junction.} 
\label{ref:scheme}
\end{figure}

In this Letter, we introduce the Josephson junction made of two NCSs weakly linked by a uniaxial ferromagnet with an easy axis normal to the interface, i.e. parallel to the electric current (see~\Figref{ref:scheme}). Unlike in previous proposals~\cite{B2008,Zhang,Rahnavard}, the ferromagnetic exchange field $\h$ here is directed normally to the NCS/F/NCS interfaces. Parity breaking in NCS couples the magnetization $\h$ to the supercurrent $\j$, resulting in a term $\j\cdot\h$ in the Ginzburg-Landau free energy functional describing crystal structure with $O$ point group symmetry. As derived below, it results in a nonzero current even in the absence of phase gradients across the junction. This current, directed along the magnetic field, stems from  the breaking of parity in a non-equilibrium state and is thus a direct analog of the Chiral Magnetic Effect \cite{ref:CME} predicted for systems of chiral fermions, and observed in Dirac and Weyl semimetals~\cite{Kharzeev2009,Li2016,Xiong2015,Huang2015}. This analogy motivates our terminology ``Chiral Magnetic Josephson junction" (CMJ junction) to describe the NCS/F/NCS junction displayed in \Figref{ref:scheme}. Below, we demonstrate that the current across the CMJ junction is still given by the expression \Eqref{current}, where the magnitude of the bias $\varphi_g$ can be tuned by the ferromagnet's magnetization.

\vskip0.3cm

We propose to use the CMJ junction as a constituent of a superconducting qubit. The junction's energy associated with the current~\Eqref{current} is 
\be\label{jen}
E(\varphi, \varphi_g) = E_J [1 - \cos(\varphi - \varphi_g)],
\ee
where $E_J$ is the Josephson energy.  The total energy of the qubit $E_Q$ is the sum of the junction's energy \Eqref{jen} and a term quadratic in the phase difference $\varphi$. For example, in the case of an inductively shunted junction \cite{Girvin}, this quadratic term results from the inductive energy $E_L$:
\be\label{eq:E:Q}
E_Q(\varphi, \varphi_g) = E_J [1 - \cos(\varphi - \varphi_g)] + E_L \varphi^2 .
\ee
Here, the offset $\varphi_g$ plays the role of an offset flux, and can be used to control the form of the qubit Hamiltonian. Using the magnetization of the ferromagnetic link should simplify the qubit architecture by avoiding the use of an offset flux, and the corresponding source. 

Noise in the offset flux is an important component of qubit decoherence~\cite{Devoret,Devoret:2}. As demonstrated below, the noise in the offset phase $\varphi_g$ results from the fluctuations of the component of the magnetization normal to the interface. In the proposed setup, this direction corresponds to the easy axis of the uniaxial ferromagnet. Thus only longitudinal fluctuations of magnetization contribute to the noise, but these are suppressed by the ratio of the qubit temperature to Curie temperature of the ferromagnet, which is about $10^{-5} - 10^{-4}$. Moreover, the current in the CMJ junction is parallel to magnetization, and thus is ``force-free", i.e. not subjected to a Lorentz force. This greatly reduces the coupling between the current and magnetization that contributes to the noise.

\vskip0.3cm

The offset phase $\varphi_g$ of the CMJ junction can be estimated within the Ginzburg-Landau (GL) framework. The superconducting state in non-centrosymmetric superconductors is commonly believed to be a mixture of singlet and triplet pseudo-spin states~\cite{Bauer.Sigrist,NCS1} due to the spin-orbital coupling in the presence of the broken inversion symmetry~\cite{ref:Gorkov}. Using $\hbar = c = 1$, the Ginzburg-Landau free energy describing the superconducting state of non-centrosymmetric material reads as \cite{ref:Agterberg,NCS1}:
\be
f & = & a |\psi|^2 + \gamma |\D\psi|^2 + \frac{b}{2} |\psi|^4 + \frac{K}{2} \j\cdot\h.
\label{eq:f:GL}
\ee
The single-component superconducting order parameter $\psi = |\psi| e^{i \varphi}$ is coupled to the vector potential $\A$ of the magnetic field $\h=\Curl\A$, via the gauge derivative $\D = - i \Grad - 2 e \A$, while the coefficients $b$, $\gamma$ and $a = \alpha (T - T_c)$ are standard phenomenological GL parameters. The parity-odd nature of the non-centrosymmetric superconductor is reflected by the last term (Lifshitz invariant) of the free energy~\Eqref{eq:f:GL}, which describes the direct coupling of the magnetic field ${\bs h}$ to the usual, parity-odd component of the supercurrent density:
\be
{\bs j} \equiv {\bs j}^{\mathrm{odd}} = 2 e \gamma \left[\psi^* {\bs D} \psi + \psi ({\bs D} \psi)^*\right].
\label{eq:j:odd}
\ee
Note that the exchange field ${\bs h}$ of the ferromagnet plays here the role of the background magnetic field ${\bs B}$.
The parity-odd, last term in \Eqref{eq:f:GL} yields an additional, parity-even, contribution
to the total supercurrent $\J$: 
\be
{\bs J} = {\bs j}^{\mathrm{odd}} + {\bs j}^{\mathrm{even}}, 
\qquad 
{\bs j}^{\mathrm{even}} = 4 e^2 \gamma K |\psi|^2 {\bs h}.
\label{eq:J:tot}
\ee
%
The GL functional \Eqref{eq:f:GL} describes NCS materials with $O$ point symmetry,  such as Li$_2$Pt$_3$B \cite{Badica.Kondo.ea:05,Sigrist} and Mo$_3$Al$_2$C \cite{Karki.Xiong.ea:10,Bauer.Rogl.ea:10}, and the coupling constant $K$ determines the magnitude of the superconducting magnetoelectric effects that follow from the broken inversion symmetry. Our derivation equally applies to non-centrosymmetric superconductors with other crystallographic groups with a generic Lifshitz invariant $K_{\alpha\beta} h_\alpha j_\beta$. In this case, when the $x$-axis is directed across the normal link, the diagonal element $K_{xx}$ should be nonzero. Notice that Lifshitz invariants of the type ${\bs n} \cdot {\bs h} \times {\bs j}$ do not have such diagonal element and thus cannot satisfy this requirement (see footnote~\footnote{Possible candidates for the NCS superconductors should thus have a crystalline structure with either the point group $O$ (Li$_2$Pt$_3$B, Mo$_3$Al$_2$C), $T$ point group (e.g. LaRhSi, LaIrSi) or $C_4$ (\ch{La}{5}\ch{B}{2}\ch{C}{6}), $C_2$ (UIr) etc. On the other hand, the point groups $C_{n\nu}$ with $n=2,3,4,6$ (possessed, for example, by the compounds Mo\ch{S}{2}, MoN, GaN, Ce\ch{Pt}{3}Si, CeRh\ch{Si}{3}, CeIr\ch{Si}{3}~\cite{NCS1}) correspond to the Lifshitz invariants of the type ${\bs n} \cdot {\bs h} \times {\bs j}$ that do not fit our proposal.
}).

As illustrated in \Figref{ref:scheme}, we consider a pair of identical non-centrosymmetric superconductors separated by a uniaxial ferromagnetic weak link whose internal exchange field $\h\equiv h_x {\bs{\mathrm{e}}}_x$ points across the link. 
We neglect the term quartic in the condensate and disregard inhomogeneities of both the condensate $\psi$ and the exchange field $\h$ in the transverse $yz$ plane.  The minimization of the GL free energy~\eq{eq:f:GL} with respect to the superconducting order parameter in the background of the ferromagnetic exchange field $\h$ then yields the equation
\be
a \psi - \gamma \frac{\partial^2 \psi}{\partial x^2} - 2 i e \gamma K h_x \frac{\partial \psi}{\partial  x} = 0
\label{eq:GL:eq}
\ee
that describes the tunneling of the Cooper pairs across the weak link. For the time being we assume the absence of an external electromagnetic field at the link, $\A=0$. 
Due to proximity effects, the tunneling of the Cooper pairs between the non-centrosymmetric superconductors through the centrosymmetric weak link will not respect the parity inversion $x \to -x$, as can be seen from Eq.~\eq{eq:GL:eq}.

The general solution of Eq.~\eq{eq:GL:eq} for the superconducting gap inside the weak link reads as:
\be
\psi(x) = C_+ e^{q_+ x} + C_- e^{q_- x},
\label{eq:psi:C}
\ee
where the wavevectors
\be
q_\pm = \pm \sqrt{\frac{a}{\gamma} - \left(e h_x K\right)^2} - i e h_x K
\label{eq:q:pm}
\ee
should have a nonzero real part so that the weak link is in a normal state, thus requiring $a > a_c = \gamma (e h_x K)^2$.
The coefficients $C_\pm$ in Eq.~\eq{eq:psi:C} are determined by the boundary conditions at the interfaces of the ferromagnetic weak link with the superconductors at $x = \pm L/2$. It is customary to make a simplification using the rigid boundary conditions~\cite{B2008,ref:Golubov} which assume the absence of a barrier at the interfaces, and imply continuity of the superconducting order parameter: 
\be
\psi(x = \pm L/2) = |\Delta| e^{\pm i \varphi/2};
\label{eq:boundaries}
\ee
here $|\Delta|$ is the absolute value of the order parameter at the superconducting leads. 
\vskip0.3cm

Using the relations Eqs.~(\ref{eq:psi:C}-\ref{eq:boundaries}), together with the definition of the total current \Eqref{eq:J:tot}, yields the phase-shifted current
relation:
\be
J = J_0 \sin\left(\varphi - \varphi_g\right) , \quad
J_0 = \frac{4 e \gamma |\Delta|^2 \sqrt{\frac{a}{\gamma} - (e h_x K)^2}}{\sinh L \sqrt{\frac{a}{\gamma} - (e h_x K)^2}}, \qquad
\label{eq:J:J}
\ee
which exhibits the offset of the phase difference given by
\be
\varphi_g = e h_x K L.
\label{eq:varphi:0}
\ee
This offset, corresponding to the broken inversion symmetry, is proportional to the strength of the magnetic interaction. In other words, the presence of the nonzero phase bias $\varphi_g \neq 0$ signals the breaking of the inversion symmetry between leftward and rightward tunneling of the Cooper pairs, and leads to a nonzero current in the ``steady state'' of the junction even if the phase difference between the superconducting leads is zero, $\varphi=0$.
\vskip0.3cm

In a long junction, $L \sqrt{a/\gamma - (e h_x K)^2} \gg 1$, the current~\eq{eq:J:J} is an exponentially small quantity due to suppression of the Cooper-pair tunneling between widely separated superconducting leads. 
The limit of a short junction in the presence of parity breaking should be taken with care. In the thermodynamic equilibrium $\varphi = \varphi_g(L)$, and the current through the junction is always zero. However, in the steady state with zero phase difference $\varphi = 0$, the electric current does not vanish, and is given by the parity--even term~\eq{eq:J:tot}: 
\be\label{eq:J:CME}
{\bs J}(\varphi=0, L\to 0) = 4 e^2 \gamma K |\Delta|^2 {\bs h}.
\ee
This current plays a crucial role in the dynamics of the chiral magnetic qubit. 
As mentioned earlier, the current~\eq{eq:J:CME} shares a striking similarity with the Chiral Magnetic Effect~\cite{ref:CME}, with $\h$ and $K$ respectively playing the roles of the external magnetic field and the source of the parity breaking.
\vskip0.3cm

Some features of the phenomenon that we discuss appear also in usual centrosymmetric $s$-wave superconductors separated by the Josephson junction made of an NCS-type ferromagnet with {\emph{tangentially}}--oriented field~${\bs h}$. Even though the underlying dynamics is quite different, the latter system is described by an equation similar to~\eq{eq:GL:eq}~\cite{B2008}. However, in our case the current flows along the magnetic field, is force-free, and thus is not subjected to the noise resulting from the transverse fluctuations of magnetization.

Similar types of Josephson junctions with the ferromagnetic exchange field $\bs h$ oriented transversally between two non-centrosymmetric superconductors
with the $C_{4v}$ point group (corresponding to interactions of the type ${\bs n} \cdot {\bs h} \times {\bs j}$) have been proposed in Refs.~\cite{Zhang,Rahnavard}. The Josephson current in a non-ferromagnetic junction between two non-centrosymmetric superconductors does not exhibit the offset in the phase~\cite{ref:Asano}. A finite phase offset for the Josephson current of chiral charge appears, however, for the junction between two Weyl superconductors separated by a Weyl semimetal, also for magnetic field oriented transversally~\cite{Zhang:2018wzm}.

\vskip0.3cm

We estimate the phase bias~\eq{eq:varphi:0} numerically as follows:
\be
\varphi_g \simeq 1.5 \times 10^{-3} \, h (\mathrm{T} ) \, L(\mathrm{nm}) \, K(\mathrm{nm}).
\label{eq:varphi:est}
\ee
The length of the ferromagnetic Josephson junction is typically of the order of tens of nanometers ($L \sim 30 \, {\mathrm{nm}}$ in Ref.~\cite{FerroJunction}). The exchange field $h$ should not exceed the upper critical field $H_{c2}$ which, for a number of NCS superconductors, may reach significant values $H_{c2} \sim 10$\,T. The fields of this order and higher are known to be created by usual ferromagnets~\cite{Buzdin:Review}. Note that even for magnetic fields larger than $H_{c1}$, the vortex formation can be avoided by choosing a weak link with a sufficiently small cross section, such that a total magnetic flux entering the superconductor is smaller than the flux quantum $\Phi_0$. The main uncertainty in our estimate comes from the poorly known parity-odd coupling $K$. Its value was estimated to be $ K \simeq (10^{-3} \dots 10^{-2}) \lambda$~\cite{Lu.Yip:09,Kashyap.Agterberg:13}, where $\lambda \simeq (0.1 \dots 1) \mu$m is the penetration depth. In spite of this uncertainty, it appears that the phase bias may be tuned to take values of order $\varphi_g \sim \pi$. For a given NCS superconductor, the phase bias can be manipulated by the magnetization of the weak link.\vskip0.3cm

\begin{figure}[htb]
\begin{center}
\includegraphics[scale=0.2,clip=true]{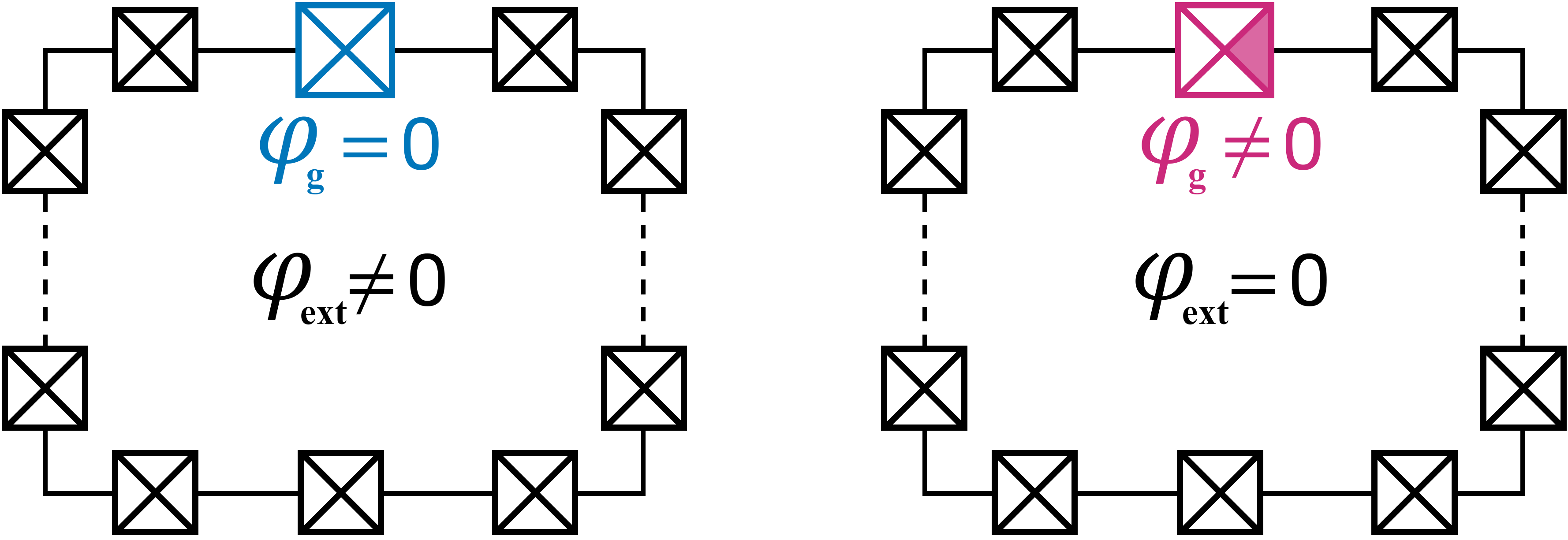} \\[1mm]
(a) \hskip 35mm (b) 
\end{center}
\vskip -5mm
\caption{(a) Fluxonium-type qubit based on conventional Josephson junction inductively shunted by a series of Josephson junctions. The qubit is biased by an external magnetic flux $\varphi_{\mathrm{ext}} \equiv 2\pi \Phi/\Phi_0$, where $\Phi_0$ is the elementary flux quantum. The gate phase offset for a conventional Josephson junction is absent, $\varphi_{{g}} = 0$. (b) Chiral magnetic qubit based on the Chiral Magnetic Josephson (CMJ) junction inductively shunted by a series of Josephson junctions. The CMJ junction possesses an internal phase offset $\varphi_{{g}} \neq 0$ eliminating the need for an external magnetic flux $\varphi_{\mathrm{ext}}$.} 
\label{ref:schemes}
\end{figure}

The chiral magnetic Josephson junction sketched in \Figref{ref:scheme} can be inductively shunted, for example by a series of conventional Josephson junctions, to form a  ``chiral magnetic qubit" (see \Figref{ref:schemes}). Such circuits include, in addition, two mixed Josephson junctions between the conventional and NCS superconductors. These mixed junctions  do not generate the electric current across them at zero phase difference $\varphi=0$~\cite{Sigrist:2}.

The Coulomb interactions between the Cooper pairs is described by the kinetic term in the Hamiltonian of the qubit:
\be\label{eq:H}
{\hat H} = 4 E_C {\hat n}^2 + E_J [1 - \cos(\varphi - \varphi_g)] + E_L \varphi^2,
\ee
where $\hat n = - i \hbar \partial_\varphi$ is the operator of the Cooper-pair number, and last two terms describe the Josephson tunneling and the induction~\eq{eq:E:Q}. 
The Hamiltonian~\eq{eq:H} is generic for a family of inductively shunted qubits including the fluxonium~\cite{ref:fluxonium:1,ref:fluxonium:2}, one-junction flux qubits, and flux-biased phase qubits~\cite{ref:Wendin}.  

As illustrated in \Figref{ref:schemes}(a), fluxonium qubits relate the phase offset to the externally applied flux $\Phi$ as $\varphi_g = 2\pi\Phi/\Phi_0$, where $\Phi_0 = h/(2 e)$ is the flux quantum. These are further characterized by a specific set of  model parameters, such as the small inductive energy ($E_L/E_J \simeq 0.045$) and a moderate charging energy ($E_C/E_J \lesssim 1$) which give a unique combination of long coherence time and large anharmonicity of the energy levels~\cite{ref:fluxonium:3}.
Transmon qubits, on the other hand, are characterized by Coulomb charging energy which is much smaller than the Josephson tunneling energy, $E_C \ll E_J$, thus allowing the reduction of noise caused by the offset charge fluctuations~\cite{ref:Koch:2007}.

\begin{figure}[htb]
\begin{center}
\includegraphics[scale=0.325,clip=true]{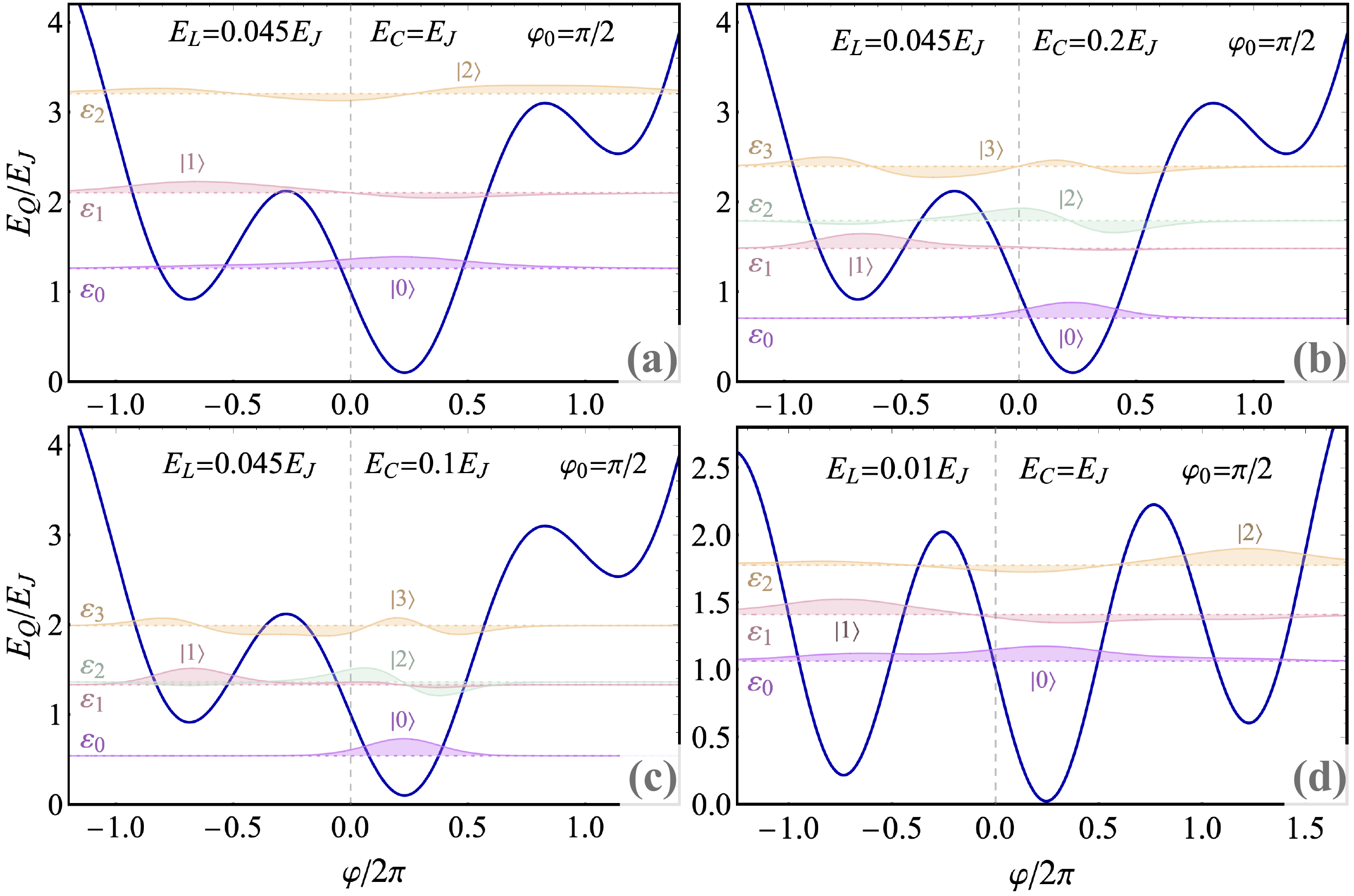}
\end{center}
\vskip -5mm
\caption{The potential energy~\eq{eq:E:Q} of the chiral magnetic qubit with the chiral magnetic Josephson junction possessing the phase offset $\varphi_g = \pi/2$ for 
various Coulomb charging energies $E_C$ and inductive energies $E_L$. Lowest eigenstates $|n\rangle$ with $n=0,1,\dots$ are shown along with the numerically computed energy levels $\varepsilon_n$ and the corresponding wavefunctions~$\psi_n(\varphi)$.} 
\label{ref:potential}
\end{figure}

\vskip0.3cm
A nonzero phase bias $\varphi_g \neq 0$ imposes a large anharmonicity on the energy-level structure~\cite{Girvin} determined by the Schr\"oringer equation:
\be
{\hat H} \psi_n(\varphi) = \varepsilon_n \psi_n(\varphi).
\ee
The regime $\varphi_g = \pi/2$ provides maximum level splitting and the absence of nearly-degenerate level pairs~\cite{Girvin}. \Figref{ref:potential} displays the structure of the energy levels corresponding to this Hamiltonian. The transitions between the first excited state and the ground state, $|1\rangle \to |0\rangle$, can be substantially suppressed by the barrier separating them. This barrier is almost absent in the typical fluxonium regime ($E_L = 0.045 E_J$ and $E_C = E_J$), as the first excited energy level $\varepsilon_1$ practically coincides with the height of the barrier, \Figref{ref:potential}(a). This conclusion is valid, to a good accuracy, for a wide range of values of the phase offset~$\varphi_g$. 
\vskip0.3cm

Decreasing the Coulomb energy towards the transmon regime leads to the appearance of the prohibitive barrier for transitions $|1\rangle \to |0\rangle$ between the different wells, \Figref{ref:potential}(b). As displayed in \Figref{ref:potential}(c), further decrease of the Coulomb energy reduces the energy difference between $|1\rangle $ and $ |0\rangle$ states. The lifetime of the first excited level may be enhanced by lowering the inductive energy $E_L$. Figure~\ref{ref:potential}(d) shows that at $E_L = 0.01 E_J$ the barrier is sufficiently high to ensure a quasi-classical protection of the first excited level. Related discussions of the energy levels can be found in Refs.~\cite{ref:fluxonium:3,ref:fluxonium:4} for fluxonium-type qubits. 
\vskip0.3cm

The conventional way to induce the phase $\varphi_g$ in fluxonium qubits is to apply a background magnetic flux $\Phi$. The noise $\delta \Phi / \Phi_0$ in magnetic flux is typically of the order of $10^{-3} - 10^{-2}$; overcoming this noise is a central problem in quantum computer design. In our case, the noise in $\varphi_g$ is due to the noise in magnetization. Indeed, Eq.~\Eqref{eq:varphi:0} yields the noise relation
\be
\left( \frac{\delta \varphi_g}{\varphi_g} \right) = e K L \left( \frac{\delta h_x}{h_x} \right).
\ee
To reduce the noise in magnetization, we propose to use a highly anisotropic uniaxial ferromagnet, e.g. of magnetoplumbite type. In the case of a uniaxial ferromagnet, the 3D rotational symmetry is explicitly broken by the symmetry of the crystalline lattice, and the only surviving symmetry is 2D rotations around the easy symmetry axis of the ferromagnet, in the basal plane perpendicular to this axis.  The fluctuations of magnetization in this case are given by the simplified form of the Landau-Lifshitz-Gilbert equation with no gyroscopic term. They correspond to the rotation of magnetization around the easy axis (which points here along the $x$ axis), with fluctuating components of magnetization $h_y$ and $h_z$, but with a fixed $h_x$ which is an integral of motion. Since the phase offset $\varphi_g$ \Eqref{eq:varphi:0} depends only on $h_x$, the transverse fluctuations of magnetization will not induce noise in this quantity. 
\vskip0.3cm

Unlike the transverse ones, the longitudinal fluctuations of magnetization (i.e. fluctuations of the magnitude of ${\bs h}$) will induce a noise in the offset phase $\varphi_g$. However, longitudinal fluctuations of the magnetization are expected to be suppressed compared to the transverse ones by a factor of $c\,T/T_C$, where $T$ is the temperature, $T_C$ is Curie temperature of the ferromagnet, and $c$ is a constant of order one. Indeed, the transverse fluctuations correspond to gapless Goldstone modes with kinetic energy $\sim T$, while the longitudinal one is massive with energy $\sim T_C$. Analysis \cite{long1,long2} of the Landau-Lifshitz-Bloch equation (including both transverse and longitudinal fluctuations of magnetization) indicates that the constant $c \simeq 2/3$. Therefore, at temperatures of the superconducting qubits that are on the order of tens of milli-Kelvin, with Curie temperatures on the order of $1,000$ K, we expect the suppression of longitudinal fluctuations by a factor of $\sim 10^{-4}-10^{-5}$. This allows to expect a suppression in the noise resulting from the offset flux of CMJ junction, as compared to external flux noise, by a significant factor of $10^{-2}$.
\vskip0.3cm

The domain structure in uniaxial ferromagnets is known to crucially depend on the anisotropy \cite{Kaczer}: in weakly anisotropic ferromagnets (Landau-Lifshitz type), there is a branching of domains close to the surface, and the magnetic flux does not leave the ferromagnet. This is not a desirable domain configuration, as the magnetic field has to penetrate the superconductor. On the other hand, in strongly anisotropic ferromagnets (Kittel type), such as magnetoplumbite, the domains do not branch, and thus the magnetic flux does escape the ferromagnet. The domain structure of a thin ferromagnetic film~\cite{Bulaevskii} may be affected by the superconducting interface; this question requires further investigation.
\vskip0.3cm

To summarize, we have introduced a Josephson junction consisting of two non-centrosymmetric superconductors connected by a uniaxial ferromagnet, and have demonstrated that it exhibits a direct analog of the Chiral Magnetic Effect. We have proposed this Chiral Magnetic Josephson junction (CMJ junction) for use as an element of a qubit with parameters tunable by the ferromagnet's magnetization. The resulting Chiral Magnetic Qubit is protected from noise caused by fluctuations in magnetization, and does not require an external magnetic flux, allowing for a simpler and more robust architecture. The main uncertainty stems from the poorly known parity--odd response of non-centrosymmetric superconductors, and we believe that these materials and the properties of their interfaces deserve further studies.


\begin{acknowledgments}

The work of D.K. was supported by the U.S. Department of Energy, Office of Nuclear Physics, under contracts DE-FG-88ER40388 and DE-AC02-98CH10886, and by the Office of Basic Energy Science under contract DE-SC-0017662. 
\end{acknowledgments}

\end{document}